\documentclass[aps]{revtex4}

\usepackage{graphicx}
\usepackage{bm}
\usepackage[usenames,dvipsnames]{color}
\usepackage{psfrag}

\begin{document}

\title{Non-equilibrium critical properties of the Ising model on product graphs}

\author{Raffaella Burioni}
\affiliation{Dipartimento di Fisica and INFN, Universit\`a  di Parma,
 Parco Area delle Scienze 7/A, I-423100 Parma, Italy.}

\author{Federico Corberi}
\affiliation {Dipartimento di Matematica ed Informatica and 
INFN, Gruppo Collegato di Salerno, and CNISM, Unit\'a di Salerno,
Universit\`a  di Salerno, 
via Ponte don Melillo, 84084 Fisciano (SA), Italy.}

\author{Alessandro Vezzani}
\affiliation{Centro S3, CNR-Istituto di Nanoscienze, Via Campi 213A, 41125 Modena Italy, and
Dipartimento di Fisica, Universit\`a  di Parma,
Parco Area delle Scienze 7/A, I-43100 Parma, Italy.}

\begin{abstract}
We study numerically the non-equilibrium critical properties of the Ising model defined on direct 
products of graphs, obtained from factor graphs  without phase transition ($T_c=0$).
On this class of product graphs, the Ising model features a finite temperature phase transition,
and we find a pattern of scaling behaviors analogous to the one known on regular lattices:
Observables take a scaling form in terms of a function $L(t)$ of time, with the meaning of 
a growing length inside which a coherent fractal structure, the critical state, 
is progressively formed.  
Computing universal quantities, such as the critical exponents and 
the limiting fluctuation-dissipation
ratio $X_\infty$, allows us to comment on the possibility to extend
universality concepts to the critical behavior on inhomogeneous substrates.
\end{abstract}

\maketitle


\section{Introduction} \label{intro}

The equilibrium physics of second order phase transitions
is quite well understood nowadays, due primarily
to the development of scaling theories and of the renormalization group.
Systems above the lower critical dimension build up
an increasing coherence length $\xi$ as a finite temperature $T_c$
is approached. 
In the neighborhood of $T_c$, $\xi$ has grown
much larger than any other characteristic length, 
and physical quantities can be expressed in scaling forms
in terms of $\xi$ only. Universal quantities, such as
the exponents entering those relations, are known
to depend only on a small set of parameters as, for instance, 
the space dimensionality $d$ and the symmetry of the order parameter. 
Other features, among which
the local geometry of the underlying lattice (e.g. triangular, cubic etc ...), 
are known to affect
only non-universal quantities, e.g. the value of $T_c$. 
Scaling and universality concepts can be extended also
to far-from-equilibrium systems approaching the critical
state kinetically, as in the prototypical case of samples
quenched from high temperature  to the critical temperature $T_c$. In this case an infinite
coherence length is built out of a length $L(t)$ which grows
indefinitely in time, and scaling, with respect to $L(t)$ in this case, is
again observed. 

This general picture of critical phenomena is well established
when the underlying lattice is homogeneous, and interesting results have been
obtained in the case of fractal structures,  exploiting their scale invariance  and their 
properties under exact decimation procedures \cite{fractals}. 
However  a general discrete structures, i.e. a graph, 
which can feature strong inhomogeneities and have no a priori symmetry,  is defined by a purely topological 
information encoded in the nearest neighborhood relations between its sites, and in this case the 
situation is still unclear. This  happens despite examples of inhomogeneous graphs, ranging from disordered materials, to percolation clusters, glasses, polymers, and bio-molecules, may be abundantly found in physics, economics, chemistry and biology \cite{3diultimoscalare}.  

A first question concerns the topological properties
playing the role of the Euclidean dimension $d$ in determining the existence
of a (finite temperature) phase transition. 
In the case of continuous symmetry models \cite{10e11divettoriali,fssgen}
(e.g. the Heisenberg model) this role has been shown to be played by 
the spectral dimension $d_s$  of the graph,
a quantity related to the low eigenvalues
behavior of the density of states in the Laplacian operator \cite{spectral}.
The spectral dimension of a graph alone determines the existence of a finite temperature phase transition, and $d_s=d_L=2$ is the lower critical dimension above which a graph can sustain
an ordered phase at finite temperature $T<T_c$. Therefore one expects a pattern of
scaling behaviors similar to the one observed on homogeneous lattices and a natural
extension of the concept of universality, with  $d_s$ playing a role analogous to the Euclidean
dimension $d$. However, the relation between the spectral dimension and the critical
exponents in the case of continuous symmetry models on a general graph is still an open problem, with a few rigorous results \cite{sferico,10e11divettoriali}.

For discrete symmetry models,  even such a simple topological generalization of the Euclidean dimension does not exists, and an indicator playing the same role as $d_s$ in this case is not known, although partial results have been obtained in this direction \cite{campari}. Discrete symmetry
models have been shown to feature a phase transition when $d_s\ge 2$ \cite{fssgen} but a necessary and sufficient condition for its existence has not been yet demonstrated, and also the effect of topology on critical exponents is still a completely open problem.  In particular, the knowledge of $d_s$ is not sufficient  to determine the universality class  of a discrete symmetry model.  
This poses a number of important questions yet to be answered about the critical properties
of discrete symmetry models on general graphs. Lacking an equivalent of $d_s$, the problem of distinguishing graphs where a phase-transition
occurs at finite $T_c$ from those where $T_c=0$, is opened. 

Let us recall that on usual lattices the lower critical dimension 
of the model is $d_L=1$, whereas a two-dimensional lattice with $T_c>0$ can
be obtained as the direct product (as defined in Sec. \ref{graphs}) of two one-dimensional systems. 
Extending this observation to the realm of inhomogeneous structure and general graphs, one is led to
consider direct products of graphs with
$T_c=0$. 

The direct product of  graphs represents a practical receipe to build
a class of inhomogeneous structures with $T_c>0$ but no a priori symmetry, allowing
the study of their critical properties as a function of topology alone.  
Following this reasoning, in this paper we study numerically the non equilibrium 
critical properties of the Ising model defined on direct 
products graphs. We determine the numerical value of $T_c$, which is finite as predicted by the analytic results \cite{fssgen} and we show 
that the whole non-equilibrium scaling behavior is analogous to that found
on usual lattices with $T>T_c$. In particular, observables as the two-site/two-time
correlation function $C_{ij}(t,s)$ can be expressed in terms of a growing length $L(t)$ inside which a coherent fractal structure,
the critical state, is progressively built.
 
The behavior of the Ising model on fractal structures with $T_c=0$, considered previously in \cite{nostriscalari}, is
very similar to that observed on the usual $1-d$ lattice.
Not only on these fractal structures the model remains disordered at any finite temperature 
but, in addition, at least a couple of universal quantities
(the response function exponent $a$ of Eq. (\ref{scalchi}), and the limiting
fluctuation-dissipation ratio $X_\infty$ of Eq. (\ref{xinfty})) take always the same
value of the $1-d$ lattice \cite{nostriscalari}. This features might be interpreted as an 
indication of a sort of universality, although in a vague and weak sense, 
since other exponents are different. This makes 
the direct products of these {\it 1d-like} graphs interesting also because, 
pushing the above presumptive universality arguments even further, 
and in analogy to what happens 
on homogeneous lattices, 
one might wonder if the {\it 2d-like} graphs obtained in this way show any universal
behavior, at least in the weak sense discussed above. 
However, the numerical results of this paper indicate that all the scaling exponents, including $a$,
and also $X_\infty$, differ among the possible product graphs and with respect to the values taken
in the homogeneous case with $d=2$. This proves that some other topological differences
are relevant (in a renormalization group sense),
making the concept of universality, if any, yet obscure. 
Understanding the nature of these relevant features,
and, possibly, the way to associate them some topological indexes, analogous 
to $d$ and $d_s$, remains an open problem.

The paper is organized as follows: In Sec. \ref{graphs} we recall some basic notions
about graphs. Sec. \ref{scaling} contains an overview of the scaling
behavior on homogeneous structures, and extensions to the realm of
the inhomogeneous ones. In Sec. \ref{results} we study numerically the
critical properties of the Ising model on some product graphs, determining the 
critical temperature and the scaling properties of correlation and response
functions.
Sec. \ref{below} contains a brief analyses of the kinetics of the model after a quench
below $T_c$, in order to comment on the issue of universality also in this case.
In Sec. \ref{concl} we draw our conclusions and discuss some open problems.

\section{Generalities on graphs} \label{graphs}

A graph (network) $\cal G$ is defined by a countable set of sites $i$ connected pairwise by unoriented links  $\{i,j\}$. The chemical distance $r_{i,j}$ \cite{grafi}, i.e. the number of links of the shortest path connecting  sites $i$ and $j$, defines a natural metric on $\cal G$.  The van Hove sphere ${\cal S}_{o,r}$ of radius $r$ and center $o$ in this metric is the sub-graph of $\cal G$ composed by the sites whose distance  from $o$ is smaller than $r+1$. We call $N_{o,r}$ the number of sites in ${\cal S}_{o,r}$. On infinite graphs the asymptotic behavior for large $r$ of $N_{o,r}$ defines the  fractal dimension:
\begin{equation}
N_{o,r} \sim r^{d_f}
\label{fractal_dim}
\end{equation}
where  $\sim$ denotes the behavior for large $r$. In the following we will consider only connected graphs embeddable in a finite dimensional space, with $d_f$ well defined and finite. We also require that the degree $z_i$ (number of neighbors of the site $i$) is bounded. On fractal graphs one can equivalently explore the infinite structure using, in a very natural way, finite generation (sub)fractals instead of the Van Hove spheres (see below for a definition of {\it generation}).

The graph can be described using a set of characteristic matrices. The adjacency matrix $A_{i,j}$ of a graph has entries equal to $1$ if $i$ 
and $j$ are neighboring sites ($\{i,j\}$  is a link) and $A_{i,j}=0$ otherwise. The Laplacian matrix $\Delta_{i,j}$ is defined as
\begin{equation}
\Delta_{i,j}=\delta_{i,j} z_{i}-A_{i,j}
\label{laplacian}
\end{equation}
where $z_i=\sum_j A_{i,j} $ is the degree of $i$. Interestingly, $\Delta_{i,j}$ is the generalization to graphs of the usual Laplacian operator of Euclidean structures \cite{rassegna}. In particular its spectrum is positive and for connected structures the constant vector is the only eigenvector of eigenvalue zero. We define the spectral density of the subgraph ${\cal S}_{o,r}$ as $\rho_r(l)=N_{o,r}^{-1}\sum_{k=1}^{N_{o,r}} \delta(l-L_{k,r})$ where $L_{k,r}$ are the eigenvalues of the Laplacian matrix of ${\cal S}_{o,r}$. Let  $\rho(l)$ be the limit of $\rho_r(l)$ for $r \to \infty$ .  For positive functions, if the thermodynamic limit $r \to \infty$  exists, then it can be shown that the result is independent of the center of the sphere $o$ \cite{rassegna}. If   $\rho(l)$  behaves for small $l$ as
\begin{equation}
\rho(l)\sim l^{d_s/2-1}
\label{spectral_dim}
\end{equation}
then $d_s$ is defined as the spectral dimension of the  graph \cite{spectral}.

Graphs of large dimension can be built by introducing the direct product. 
Given two graphs $\cal G$ and $\cal H$, the direct product ${\cal G}\times {\cal H}$ is a graph whose 
sites are labeled by a pair $(i,j)$ with $i$ and $j$ belonging to $\cal G$ and $\cal H$ 
respectively. $(i,j)$ and $(i',j')$ are neighbor sites in ${\cal G}\times {\cal H}$ 
if $i=i'$ and $\{j,j'\}$ is a link of $\cal H$, or if $j=j'$ and $\{i,i'\}$ is a link of 
$\cal G$. Interestingly, a basic properties of ${\cal G}\times {\cal H}$ is that, calling  $d_s^{\cal G}$ 
and $d_f^{\cal G}$ the dimensions of the graph $\cal {G}$, the following relations hold \cite{rassegna}
\begin{equation}
d_s^{{\cal G} \times {\cal H}} = d_s^{{\cal G}}+d_s^{{\cal H}}~~~~~~~
d_f^{{\cal G} \times {\cal H}} = d_f^{{\cal G}}+d_f^{{\cal H}}~~~~~~~
\label{sum_spectral_dim}
\end{equation}
i.e. the spectral and fractal dimensions of the product graph are the sum of the dimensions of the original graphs.

\begin{figure}
\begin{centering}
\includegraphics[width=8.5 cm]{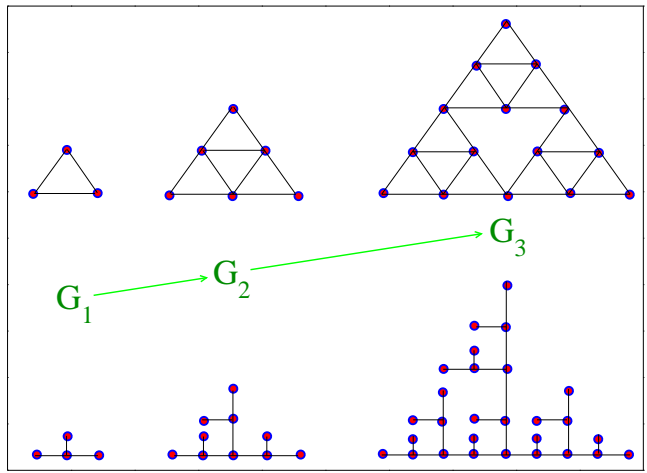}
\caption{(Color online.) Recursive construction of the SG and of the TF.}
\label{figgener}
\end{centering}
\end{figure}

We will consider models defined on graphs with known $d_s$ and $d_f$ 
to verify the relevance of these dimensions on universal properties 
and we will use direct products to obtain graphs with a finite critical temperature.
In particular, we use as base graphs two finitely ramified fractals \cite{aharony} with $d_s<2$, 
the T-fractal (TF) and the Sierpinski gasket (SG).
These fractals can be built recursively as shown in Fig. \ref{figgener}.
For the SG one starts with three spins located on the vertexes of a triangle. This is the
first generation $G_1$ of the structure. 
The second generation $G_2$ is obtained by
considering three adjacent $G_1$ structures, building a larger triangular
object. The process is then iterated at will. A similar construction
is used for the TF, as shown in Fig. \ref{figgener}, the first generation is $G_1$ and
the generation $G_n$ can be built attaching 3 $G_{n-1}$ structures in a single site. 
We will denote with $R_n$ the size of the finite generation $G_n$ i.e. the maximum distance 
between two sites of $G_n$.
The fractal and spectral dimensions of the TF and of the SG can be analytically evaluated 
by means of exact renormalizations \cite{rammal}, yielding 
$d_f=\log(3)/\log(2)$, $d_s=\log(9)/\log(6)$ and $d_f=\log(3)/\log(2)$, $d_s=\log(9)/\log(5)$ 
respectively. We  consider the graphs obtained from the product of two SG's (SGxSG) and 
two TF's (TFxTF), their fractal and spectral dimensions can be calculated from 
Equations (\ref{sum_spectral_dim}) obtaining 
$d_f=3.17..$, $d_s=2.45...$ for the TFxTF and $d_f=3.17..$, $d_s=2.73...$ for the SGxSG. 
Since $d_s>2$ the Ising model present a  phase transition at finite temperature $T_c$, 
according to the generalized Froelich-Simon-Spencer bound \cite{fssgen}. 

\section{Models and Scaling Relation} \label{scaling}

In this section we introduce the model, the dynamical quantities of interest and overview their 
known scaling behavior.  We consider, on a given graph ${\cal G}$, the Ising model, defined by the Hamiltonian
\begin{equation}
H[\sigma ]=-J\sum _{<ij>}\sigma _i \sigma _j \equiv  -J\sum _{i,j}A_{i,j}\sigma _i \sigma _j 
\label{hamiltonian}
\end{equation}
where $\sigma _i =\pm 1$ denotes the spin of site $i$, $<ij>$ are
nearest neighbors on the graph and $A_{i,j}$ is the adjacency matrix. The dynamics is introduced by randomly choosing a single spin 
and updating it with Metropolis transition rates:
\begin{equation}
w([\sigma]\to [\sigma'])=\min \left [1, \exp (-\Delta E/(K_BT))\right ].
\label{metropolis} 
\end{equation}
Here $[\sigma]$ and $[\sigma']$ are the spin configurations before and
after the move, $K_B$ is the Boltzmann constant, and
\begin{equation}
\Delta E=H[\sigma']-H[\sigma].
\label{deltaE}
\end{equation}
A Montecarlo time step is a sequence of $N$ random moves where $N$ is the number of 
sites in the finite realization of the graph $\cal{G}$. 
We will focus on the non-equilibrium situation where a completely random 
initial configuration, corresponding to an equilibrium state at
$T=\infty$, is instantaneously quenched to a 
temperature equal or smaller than $T_c$.

\subsection{Quenches to $T_c$} \label{scalat}

After quenching our system to the critical temperature,
the two-time correlation function
in the case of Ising variables is defined as
\begin{equation}
C_{ij}(t,s)=\langle \sigma _i(t)\sigma _j(s)\rangle 
\label{gencorr}
\end{equation}
($\langle \ldots \rangle $ being an average over initial conditions and thermal
histories). On a homogeneous structure (i.e. a lattice) for sufficiently large times 
(\ref{gencorr}) takes the scaling form \cite{godluck,cal}
\begin{equation}
C_{ij}(t,s)=L(s)^{-bz} g\left[\frac{r}{L(s)},\frac{L(t)}{L(s)}\right] 
\label{genscal}
\end{equation}
where $L(t)\propto t^{1/z}$ is the typical length associated to the growth of the critical phase, $z$ is the usual equilibrium dynamic exponent
relating the relaxation time to the coherence length $\tau \propto \xi ^z$,
$r$ is the Euclidean distance between sites $i$ and $j$, $g(x,y)$ is a scaling function, 
and $bz=d-2+\eta$ is an exponent
related to the usual equilibrium critical one $\eta$. 
The scaling form (\ref{genscal}) is determined by the growth of
correlated regions of size $L(t)$ with a fractal dimension $D_f=d-zb/2$ \cite{coniglio}.
Notice that when $d\to d_L$, $T_c\to 0$ and $b\to 0$. Then $D_f\to d$ 
and these regions grow compact. For instance, for the Ising model we are 
considering here one has $\eta =1$ at $d=d_L=1$, leading to $b=0$.
This is a manifestation of the fact
that the quench at $d_L$ is not a {\it critical} quench, but rather
resembles a quench in the ordered phase, as it will 
be discussed in Sec. \ref{scalbelow}. 
Let us remark that the analysis of the out of equilibrium process
provides the equilibrium static and dynamic exponents, since they
enter the form (\ref{genscal}). 

In the case of inhomogeneous structures, one might still expect some sort of scaling to hold. However, in that case a unique definition of distance $r$ is not available. Hence, in what follows, it will be useful to introduce the 
space integrated correlation function
\begin{equation}
F_{G}(t)=\sum _{i,j\in {G}} C_{ij}(t,t),
\end{equation}
where $G$ is a certain subset of sites, which has the advantage of
a straightforward generalization to fractal
structures. Notice that we have restricted the definition to 
the equal time correlation $s=t$, since only this function will be considered in Sec. \ref{results}.
In the case of homogeneous lattices, considering a box $G$ 
of size $R$, using Eq. (\ref{genscal}) and $L(t)\propto t^{1/z}$, one finds
\begin{equation}
F_{G}(t)=t^{-b+d/z} f\left[\frac{R}{t^{1/z}}\right].
\label{Fscal}
\end{equation}
On an inhomogeneous structure such as the TFxTF or the SGxSG 
we will consider the quantity  
\begin{equation}
F_{G_k}(t)=\sum _{i,j\in {G_k}} C_{ij}(t,t),
\label{Ffrac}
\end{equation} 
where sites are summed over the internal sites of the $k$-th generation $G_k$,
for which we expect the scaling
\begin{equation}
F_{G_k}(t)=t^{-b+d_f/z} f\left[\frac{R_k}{t^{1/z}}\right].
\label{Fscalfrac}
\end{equation}
On fractals the scaling hypothesis is hence naturally defined by Eq. (\ref{Fscalfrac})
where the scaling function $f$ depends on ${R_k}/{t^{1/z}}$ only.
Notice that the use of Eq. (\ref{Fscalfrac}) avoids the problem of the definition
of a distance, since the length $R_k$ is naturally associated to the
$k$-th generation. 

In Sec. (\ref{results}) we will study in detail the so called autocorrelation
function, obtained by letting $i=j$ in Eq. (\ref{gencorr}). We do this not only because
it is very often considered in aging systems, but also because,
being an on-site quantity, it circumvents the definition of distances.
In homogeneous systems this quantity does not depend on position and, 
denoting it as $C(t,s)$, from Eq. (\ref{genscal}) and $L(t)\propto t^{1/z}$ one has the scaling behavior
\begin{equation}
C(t,s)=s^{-b} h\left(\frac{t}{s}\right), 
\label{scalc}
\end{equation} 
with $h(t/s)=g[0,L(t)/L(s)]$. 
In equilibrium conditions, $C(t,s)$ is associated to the autoresponse function
$R(t,s)=\left . \delta \langle \sigma _i\rangle /\delta h_i(s)\right \vert _{h_i=0}$, 
describing the effect of an impulsive perturbing magnetic field $h_i(s)$
switched on at time $s$ on site $i$, by the fluctuation-dissipation theorem $TR(t,s)=-dC(t,s)/ds$.
When the system is out of equilibrium after a quench to $T_c$, the fluctuation-dissipation 
theorem no longer holds in general.
However, in the short time difference regime, $t/s\simeq 1$,
due to local equilibrium, observables such as $C$ and $R$ behave as
in equilibrium \cite{noiteff,noiegamba}. The constraint imposed by
the fluctuation-dissipation theorem,
then, determines the
value $a=b$ of the exponent entering the scaling form \cite{godluck,cal,noiteff,noiegamba} of $R$
\begin{equation}
R(t,s)=s^{-(a+1)} \hat h\left(\frac{t}{s}\right).
\label{scalchi}
\end{equation} 
With the scalings (\ref{scalc},\ref{scalchi}) and $a=b$, the fluctuation-dissipation ratio
\begin{equation}
X\left( \frac{t}{s}\right)= T\frac{R(t,s)}{\frac{\partial C(t,s)}{\partial s}} 
\label{flucdisratio}
\end{equation}
is a function of $t/s$.
Its limiting value 
\begin{equation}
X_\infty=\lim _{s\to \infty} \lim _{t\to \infty}X\left(\frac{t}{s}\right) 
\label{xinfty}   
\end{equation}
is of a particular interest, since it was shown \cite{godluck,cal} to 
be an universal quantity. For the Ising model in $d=2$ one has 
$X_\infty \simeq 0.33$ \cite{prendidacorbandolo}.
On inhomogeneous structures, since the autocorrelation function and the autoresponse
may be site dependent, we define them as $C(t,s)=(1/N)\sum _i C_{ii}(t,s)$ and similarly for $R(t,s)$, where $N$ is the number of sites of the finite realization of the infinite graph $\cal{G}$. 

\subsection{Quenches below $T_c$} \label{scalbelow}

When the quench is performed below $T_c$ the nature of the process changes,
because the target equilibrium state is ordered (magnetized) and degenerate.
In this case domains of the two possible equilibrium phases grow, and
their geometry is compact. The observable quantities introduced above 
split into two contributions \cite{bouchaudealtrienoi}
\begin{equation}
C(t,s)=C_{st}(t-s)+C_{ag}(t,s),
\label{split}
\end{equation}
and similarly for the other ones.
The first stationary term is a quasi-equilibrium contribution
provided by the interior of the domains, which is in fact equilibrated
in one of the two possible phases. What is left over, the non-equilibrium
character due to presence of the interfaces, contributes with the second
term, which takes the scaling form (\ref{scalc}) with $b=0$ 
(due to the compact geometry of the domains and the above-mentioned relation between
$b$ and $D_f$). Similarly, for the aging
part of the response function one has scaling as in Eq. (\ref{scalchi}).
At variance with critical quenches, there is no known relation between 
the exponent $a$ and $b$, nor with any other equilibrium or non-equilibrium exponent. 
In two dimensions the value $a=1/4$ has been conjectured, and
numerical simulations tend to conform to this hypotheses \cite{noi}
(although the value $a=1/2$ has been also reported \cite{henkel}). 
It is also known that $a>0$ for $d>d_L$, and that $a\to 0$ when $d\to d_L$.
Hence, the value of the exponents $a$ and $b$, when $d$ is lowered 
towards $d_L$ approach the value 
$a=b=0$ both along the path 
of critical quenches (at $T=T_c$) or along the route with $T=0$
(or any route with $T<T_c$).   
However, as explained in \cite{noiteff}
the quench at $d=d_L, T=0$ cannot be regarded as {\it critical}, but rather
as a quench in the ordered phase, as it is clear since the equilibrium
state at $T=0$ is degenerate and magnetized.  
Finally, let us recall that $a>0$ and $b=0$ implies that 
$X_\infty=0$ through Eqs.(\ref{flucdisratio},\ref{xinfty}).

\subsection{What is known about $a$ and $X_\infty$ on inhomogeneous structures}

We have already discussed the fact that on Euclidean geometries 
one always finds $a>0$ in the phase-ordering 
kinetics of systems above $d_L$, while $a=0$ at $d_L$ \cite{noiteff,lipzan}. 
Elaborating on this, in \cite{nostriscalari,noivec} it was conjectured that
a similar property holds also on inhomogeneous structures, and
the exponent $a$ was proposed to infer the presence/absence of
a finite temperature transition on a graph.
Indeed $a=0$ was found on 
structures such as the SG, the TF and others where a magnetized state 
cannot be sustained at any finite temperature, whereas $a>0$ was observed
for instance on the toblerone lattice (the product graph between 
the SG and the line) where $T_c>0$.
Regarding the limiting fluctuation-dissipation ratio, the data
reported in \cite{nostriscalari} yield results very well compatible  
with the value $X_\infty=1/2$ observed in the $d=1$ homogeneous case \cite{nota},
on all the structures with $a=0$ considered. Apart from $a$ and $X_\infty$,
other exponents such as $z$ were found to be temperature dependent.
In conclusion, all the graphs with
$T_c=0$ considered in \cite{nostriscalari} share the same value for two 
universal quantities, $a$ and $X_\infty$, with the usual 1d lattice.
 
\section{Numerical results: Quench at $T_c$} \label{results}

In the following we will study the critical behavior of the product
graphs TFxTF and SGxSG. In the simulations we used finite fractals of generation 7 so that 
the total number of sites equals 4787344 for the TFxTF and  1199025 for the SGxSG. We verified that the graph size is large enough to avoid finite size 
effects so that average quantities such as $C(t,s)$ turn out to be independent 
of the size $N$.
In the following we will set $K_B=1$ and $J=1$. We will consider 
first the scaling of the equal time
correlation $F_{G_k}(t)$, which, as a byproduct, allows one to
determine $T_c$, and then the properties of two-time quantities. 

\subsection{Determination of $T_c$ and scaling of $F_{G_k}(t)$} \label{firstmethod}

\begin{figure}
\begin{centering}
\includegraphics[width=8.5 cm]{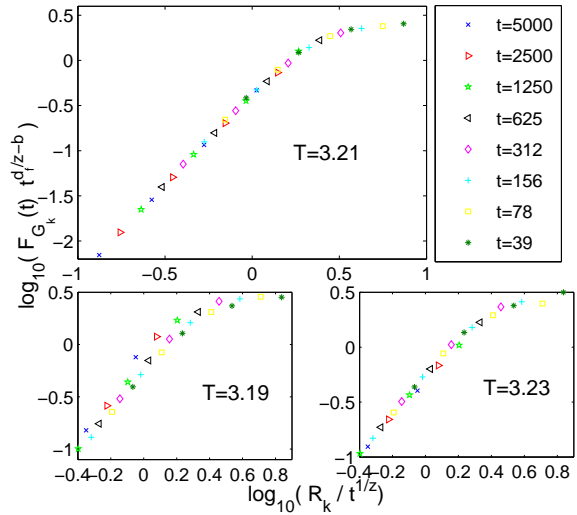}
\caption{(Color online.) Data collapse obtained by plotting $t^{b-d_f/z}F_G(t)$
against $R_k/t^{1/z}$ for the TFxTF at three temperatures, $T=3.21\simeq T_c$ (upper panel), $T=3.19$ (lower left panel) and $T=3.23$ (lower right panel). Different points are 
evaluated on TFxTF of generation $k$ with $k$ ranging from 3 to 7. The value of $S(T,z,b)$ 
is $6 \cdot 10^{-4}$ at $T=3.21$, $3 \cdot 10^{-3}$ at $T=3.19$ and $2 \cdot 10^{-3}$ at $T=3.23$}
\label{compat}
\end{centering}
\end{figure}

In order to determine $T_c$ we have followed two methods.
The first makes use of the critical finite-size scaling of $F_{G_k}(t)$,
Eq. (\ref{Fscalfrac}). This form holds right at $T_c$, with $b>0$.
Instead, above $T_c$ there is a departure from scaling at
large times, since the system eventually equilibrates.
Below $T_c$, a similar scaling form holds but only for
the aging part of the correlation (recalling the discussion
around Eq. (\ref{split})), and with the value $b=0$ characteristic
of phase-ordering. In conclusion one expects to observe deviations
from the scaling (\ref{Fscalfrac}) with $b>0$ as one moves away from $T_c$. 
Hence, the method we use for determining $T_c$ amounts to the computation of $F_{G_k}(t)$ 
at several
temperatures, trying to collapse the data according to Eq. (\ref{Fscalfrac})
by plotting $t^{b-d_f/z}F_{G_k}(t)$ against $R_k/t^{1/z}$,
using $T$, $b>0$ and $z$ are fitting parameters. More precisely,
using the least square method, we introduce a quantity 
$S(T,z,b)$ with the meaning of a variance around the optimal value 
(see the Appendix for a precise definition).
$S(T,z,b)$  measures the {\it quality} of the collapse and then we 
determine $T_c$, $z$ and $b$
as the values that provide the minimum of $S(T,z,b)$.
Examples of data collapsed in this way are shown in 
Fig. \ref{compat} for the TFxTF. In the three panels we show the behavior
of the curves collapsed with the optimal choice of $T_c$, $z$, $b$ (upper panel)
or with other choices of $T$ (larger and smaller than the optimal one,
which turns out to be $T_c=3.21$)
for which $S(T,z,b)$ is slightly
larger than the minimum. It is clearly observed that the collapse 
is better in the first case.

\begin{figure}
\begin{centering}
\includegraphics[width=11 cm]{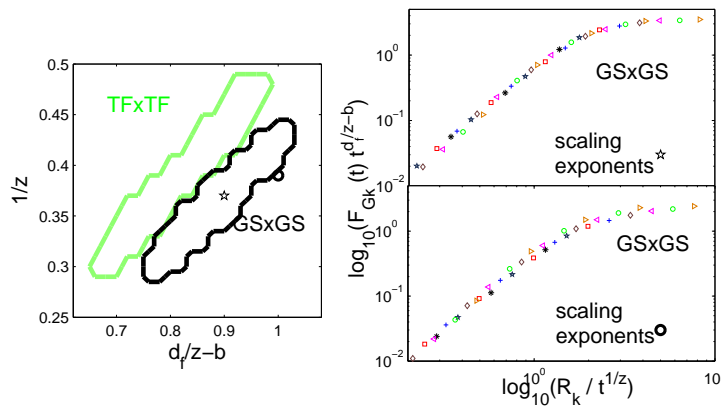}
\caption{(Color online.) In the left panel, continuous lines 
enclose the parameter compatibility regions (see text) for the TFxTF and
the SGxSG. 
The right panel represents the data collapse for $F_{G_k}(t)$ on the SGXSG. Upper plot
is obtained by using as scaling exponents the value of the center of the 
compatibility region, evidencing a very good data collapse. The lower panel refers to 
values of the exponents at the border of the compatibility region. Notice that already for $n=3$ the scaling is sensitively worse, this evidence the reliability of the definition of the compatibility region.}
\label{figtc}
\end{centering}
\end{figure}

The next point is to give an estimate of the errors.
It must be stressed here that statistical errors are quite small in our
simulations, whereas the major source of inaccuracy is a systematic
effect due to having a finite  window of times $t$ and of sizes $R_k$.
Therefore, in order to estimate the errors on the fitting parameters
$T_c$, $z$, $b$,
we define a compatibility region, namely a region around the optimal value 
of the parameters where the value of $S(T,z,b)$ is not
larger than $n$ times the minimum. Here $n$ is the tolerance
that we fix to $n=3$ (this value was chosen arbitrarily, but we have checked that 
different choices, ranging from $n=2$ to $n=5$, do not change the 
compatibility region significatively). An example of this procedure
is depicted in Fig. \ref{figtc} (left panel). Here, since we have three parameters,
for the purpose of visualization we have fixed one of them, 
the temperature, to its optimal value and we have plotted the projection
of the compatibility regions (for the TFxTF and SGxSG) on the $z,b$ plane. 
The right panels show that the collapse of the curves for the SGxSG gets worst
when one moves from the optimal value of the parameters, in the centre
of the compatibility region, to the border.
From the determination of the compatibility region we can eventually
associate an error to the determination of $T_c$, $z$, $b$ as the distance
between the optimal value and the border of the region.
This whole procedure provides the following values  
$T_c=3.21\pm0.02$, $b=0.4 \pm 0.15$, $z=2.6 \pm 0.6$ for the TFxTF, and 
$T_c=6.14\pm0.02$, $b=0.3 \pm 0.15$, 
$z=2.8 \pm 0.6$ for the SGxSG. The errors on $z$ and $b$ are quite large, therefore within 
this approach it is not clear if the critical behavior is characterized by 
the same exponents for our different structures. However Fig. \ref{figtc} 
evidences that it is very unlikely that both exponents are equal because the
compatibility regions do not intersect.

Comparing the values of $T_c$ for the TFxTF  and the SGxSG 
with the values of the lattice in two, three and four dimensions (i.e. $T_c\simeq 2.27$,
$T_c\simeq 4.51$ and $T_c\simeq 6.68$) one observes that $T_c$ tends to increase with 
the average coordination number of the graph. However, clearly other non-universal parameters play a relevant role in 
determining the critical temperature, as can be noticed from the fact that 
the SGxSG and the 4-dimensional lattice have 
both $z_i=8$ but a different value of $T_c$.

The second method for the determination of $T_c$ makes use of the scaling
properties of $C(t,s)$, as proposed and discussed in \cite{sarralip}.
With this technique we obtained results in agreement with the first
method. Let us stress the advantage of these non-equilibrium methods,
since there is no need to equilibrate the system which, due to the
critical slowing down, is a very demanding numerical task.

\subsection{Scaling of two-time quantities}

\begin{figure}
\begin{centering}
\includegraphics[width=8.5 cm]{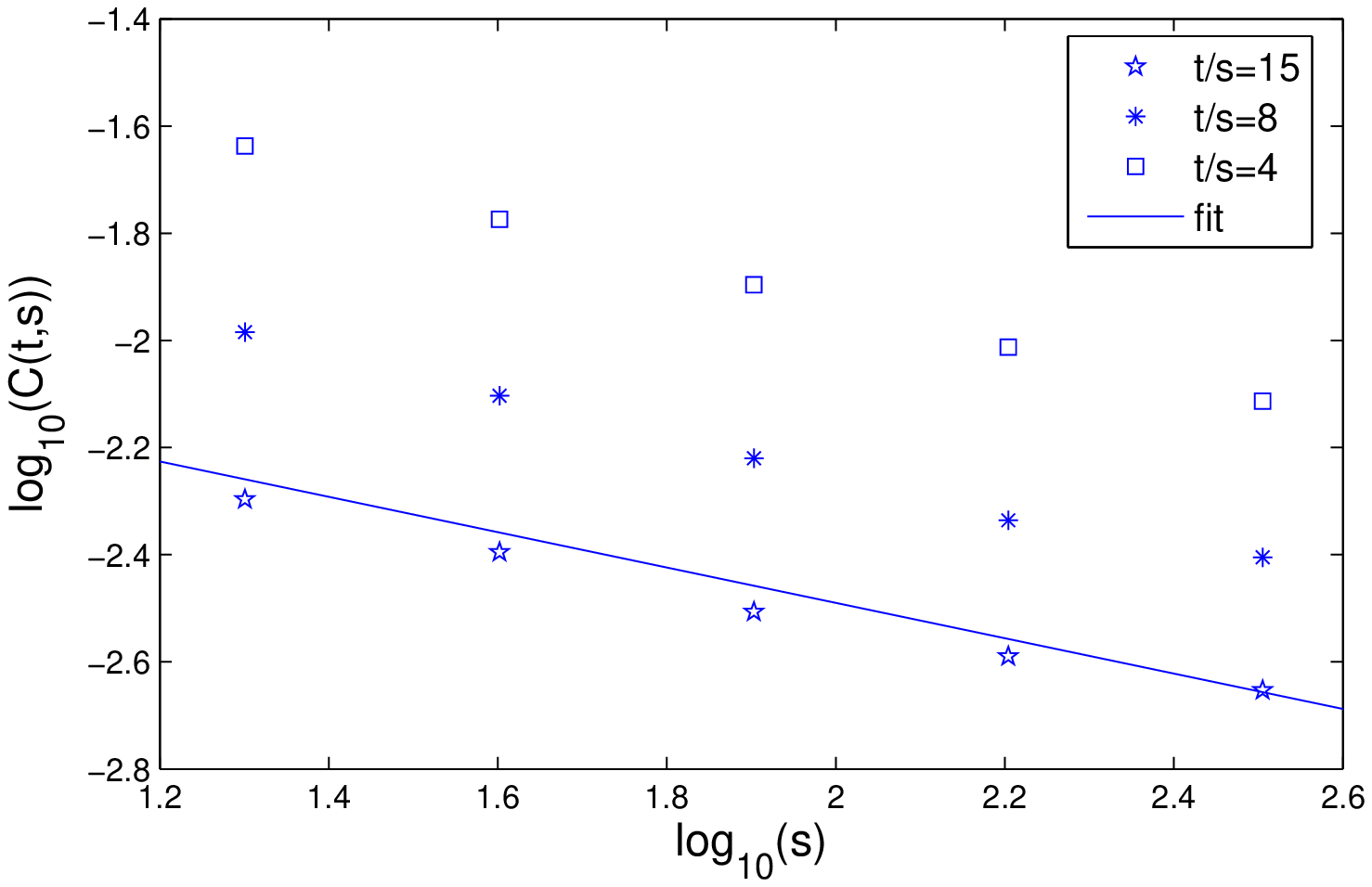}
\includegraphics[width=8.5 cm]{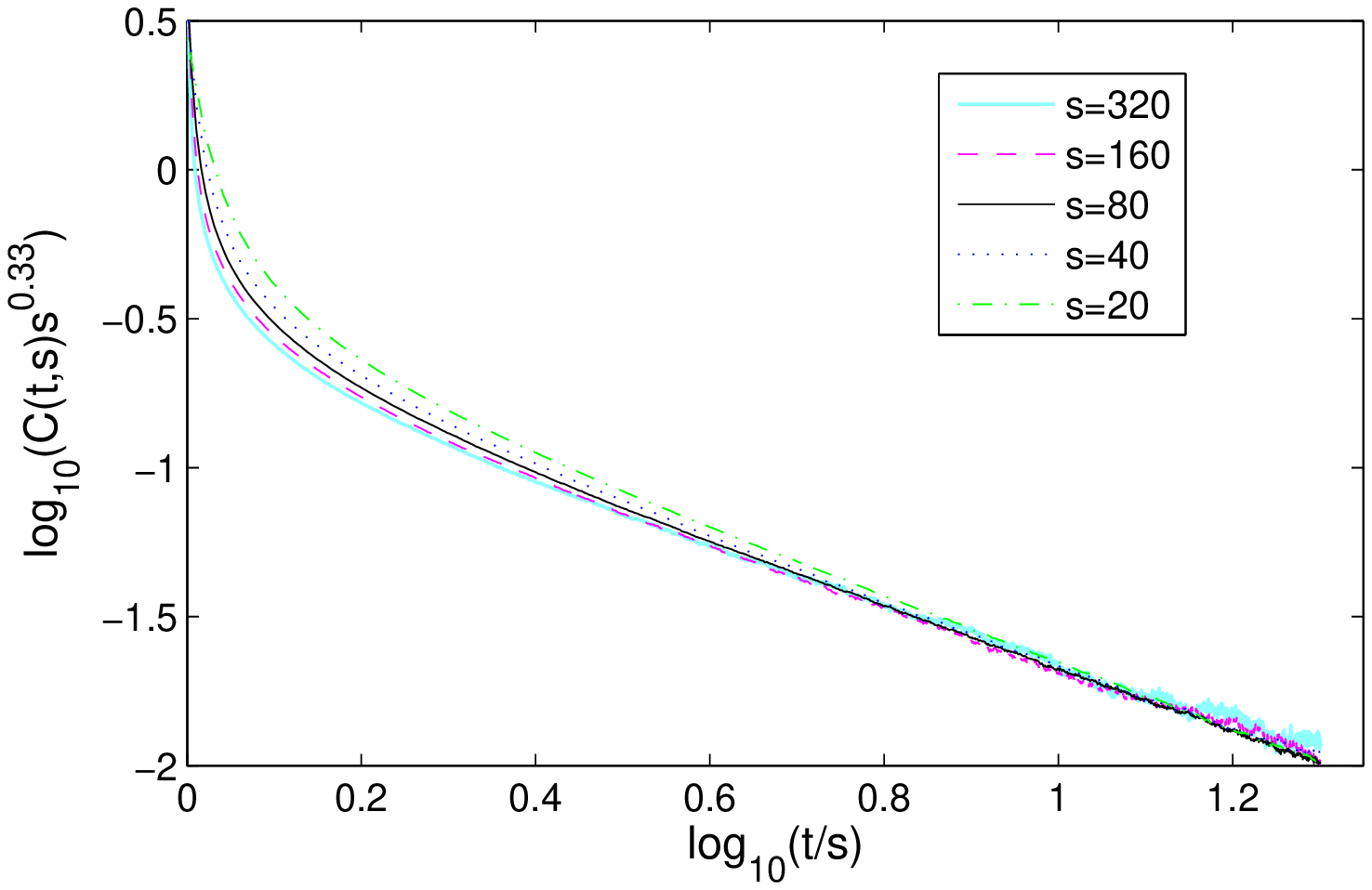}
\caption{(Color online.) Left panel: Plot of $C(t,s)$ against $s$ for different fixed values 
of $t/s$ for the TFxTF quenched to $T_c$. The continuous line is the power law $s^{-b}$ with $b=0.33$. Right panel: Data collapse obtained by plotting $s^bC(t,s)$ against $t/s$,
for several values of $s$.}
\label{figscalc}
\end{centering}
\end{figure}

\begin{figure}
\begin{centering}
\includegraphics[width=8.5 cm]{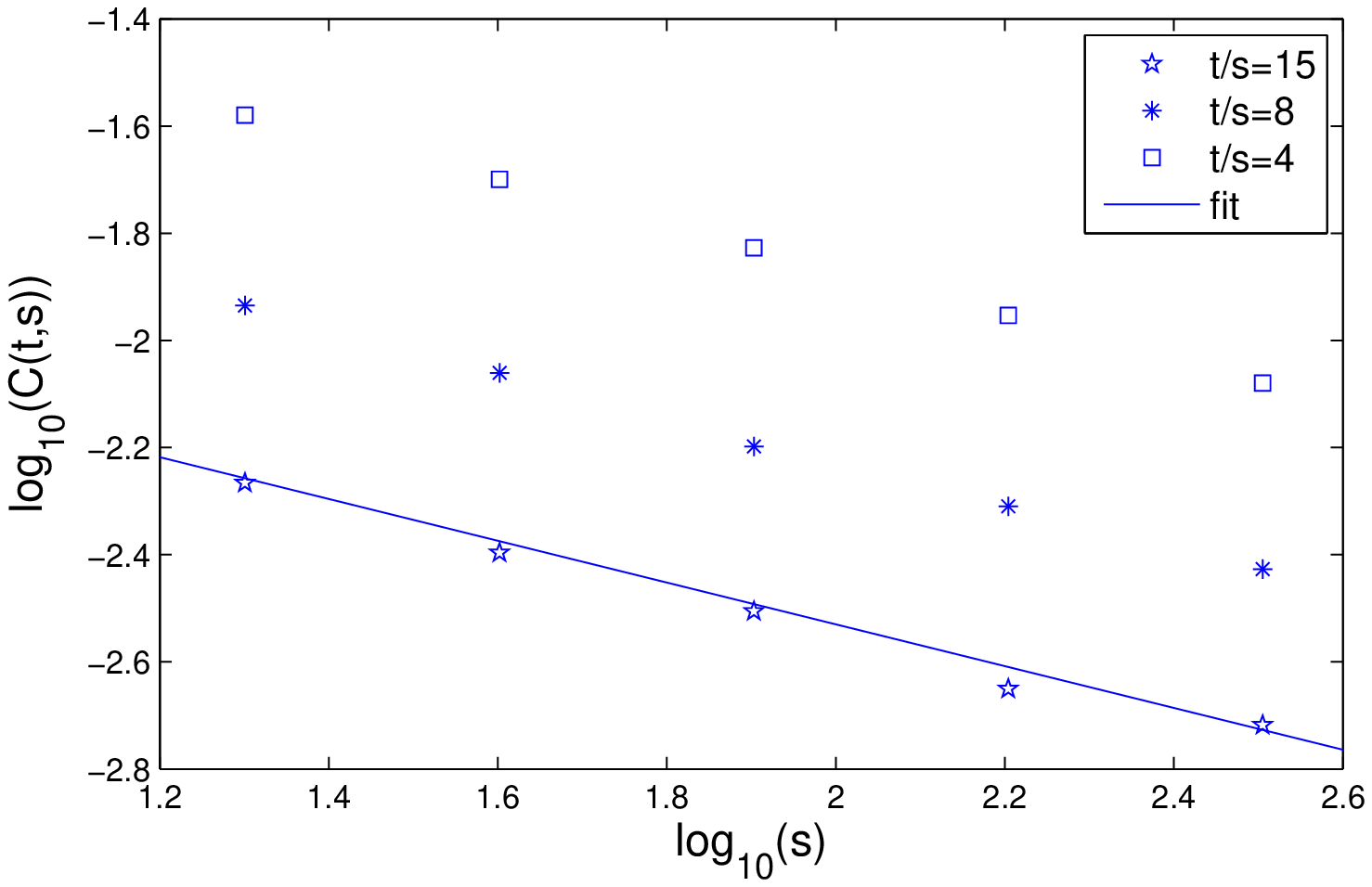}
\includegraphics[width=8.5 cm]{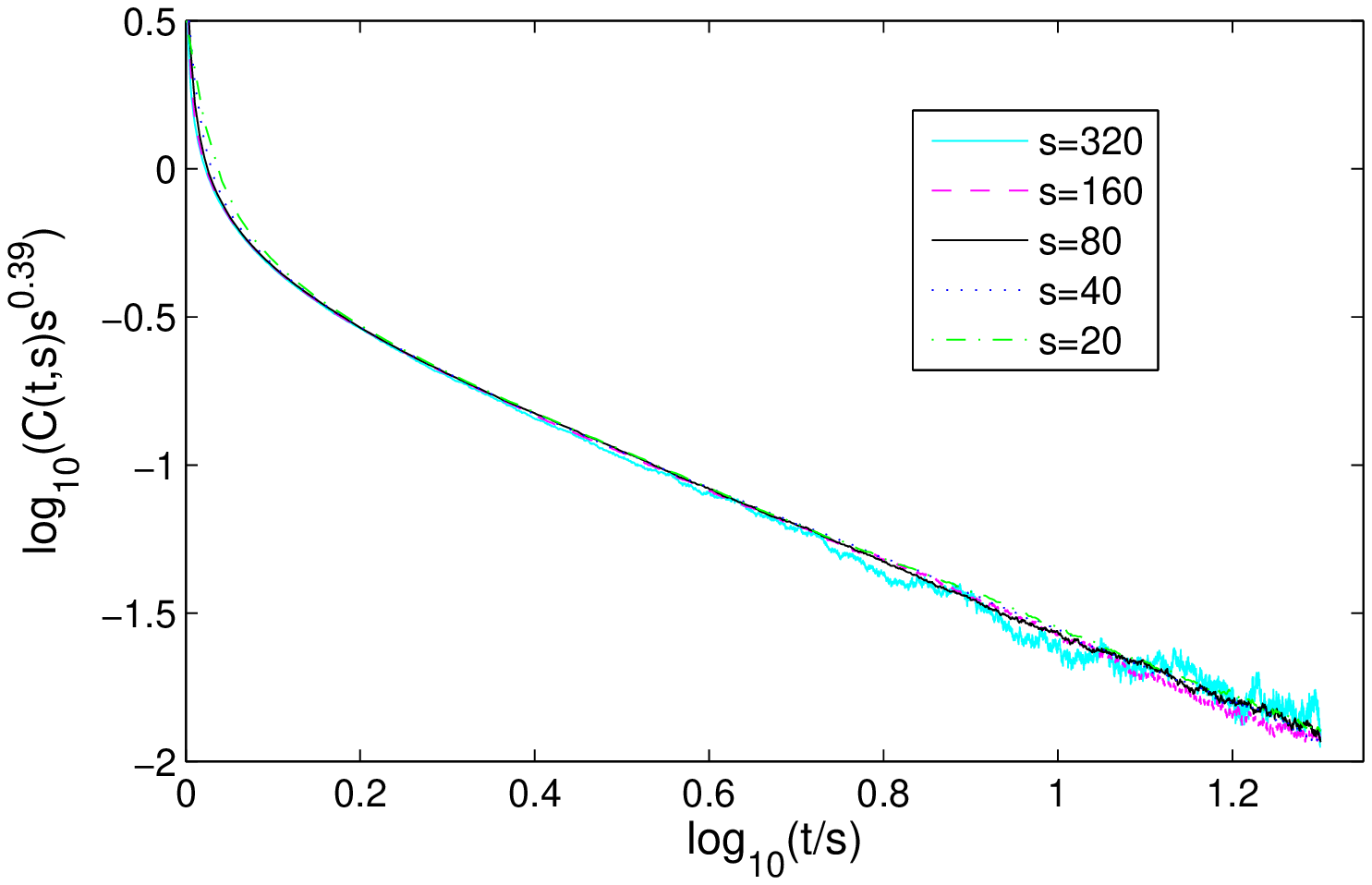}
\caption{(Color online.) Left panel: Plot of $C(t,s)$ against $s$ for different fixed values of $t/s$ for the SGxSG quenched to $T_c$. The continuous line is the power law $s^{-b}$ with $b=0.39$. Right panel: Data collapse obtained by plotting $s^bC(t,s)$ against $t/s$,
for several values of $s$.}
\label{figscalcSG}
\end{centering}
\end{figure}

Letting $T=T_c$, we have computed the two-time quantities $C(t,s)$ and $R(t,s)$.
In view of Eq. (\ref{scalc}), by plotting $C(t,s)$ for fixed values of $t/s$ against $s$, 
(left panel of Fig. \ref{figscalc} and Fig. \ref{figscalcSG}) we have determined the value $b=0.33\pm 0.03$ for the TFxTF 
and $b=0.39\pm0.03$ for the SGxSG, in good agreement with the previous determination obtained 
through $F_G(t)$. Notice that the two time approach provides a much more 
precise value of $b$: indeed a single scaling exponent 
have to be fitted while in the scaling approach to $F_{G_k}(t)$ both the exponents $b$ and $1/z$
are evaluated numerically.
Reconsidering the results of Sec. \ref{firstmethod},
we obtain also a better estimate of the $z$'s, 
which turn out to be $z=2.9\pm0.3$ and $z=2.4\pm0.2$ for the TFxTF and SGxSG respectively. 
Notice that these values (particularly $b$) are very different from those known for the 
homogeneous 2d lattice, namely $b\simeq 0.115$ and $z\simeq2.17$.
With a good confidence one may also conclude that they are
different also for the two product graphs (although, in principle,
the values $b=0.36$, $z=2.2$, obtained at the upper/lower limit of the
error bars could be compatible with both the structures).    
The data collapse
obtained by plotting $s^{b}C(t,s)$ against $t/s$ can be checked in
the right panel of Fig. \ref{figscalc}  and \ref{figscalcSG}. For large values of $t/s$ the collapse
is very good. Deviations are observed in the small $t/s$-region, but these appear to
become less important as $s$ increases. This is a quite common feature in phase-ordering,
observed also on homogeneous lattices, and can be attributed to pre-asymptotic corrections 
\cite{noiegamba,noi,preas}.

\begin{figure}
\begin{centering}
\includegraphics[width=8.5 cm]{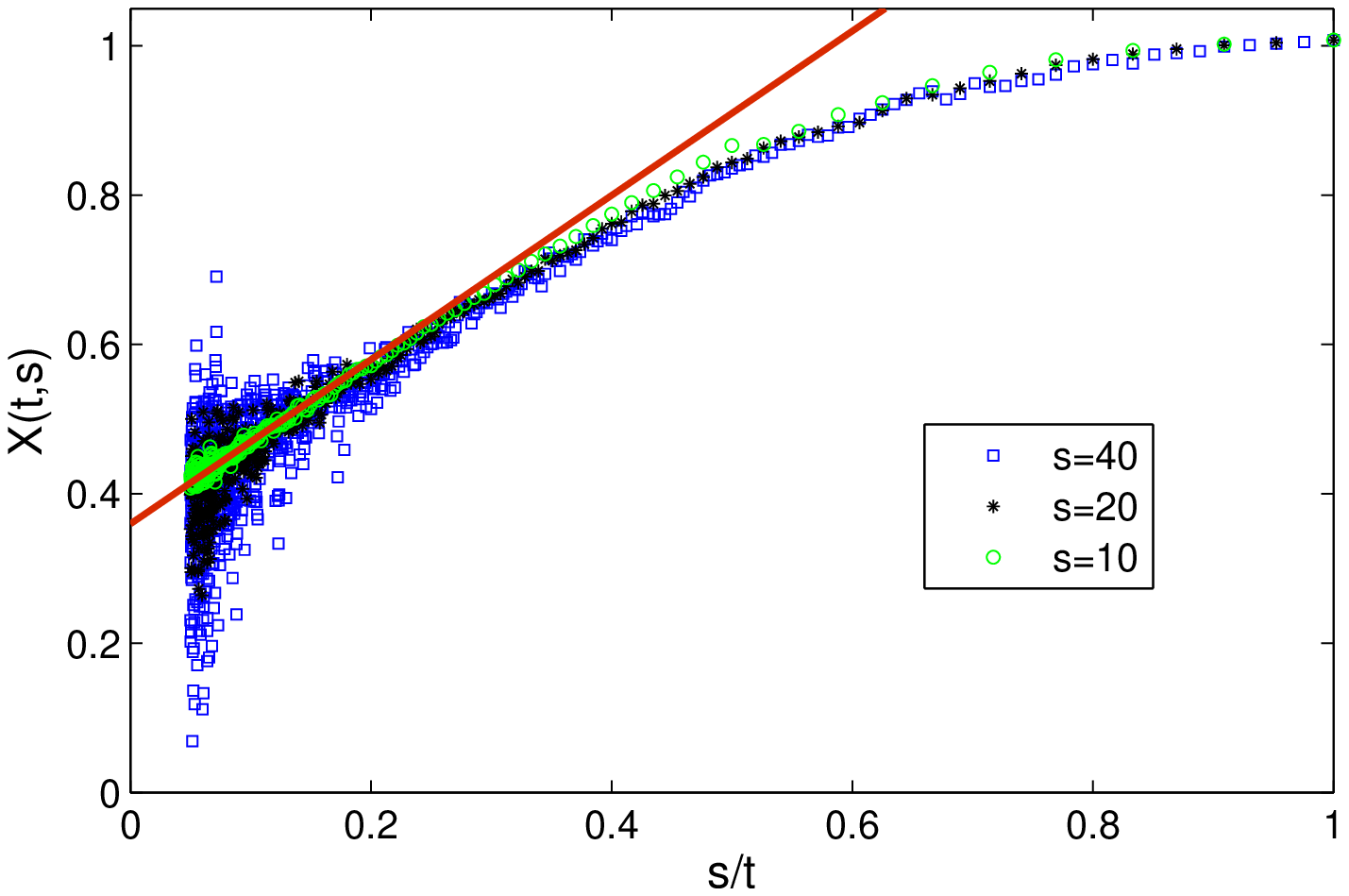}
\includegraphics[width=8.5 cm]{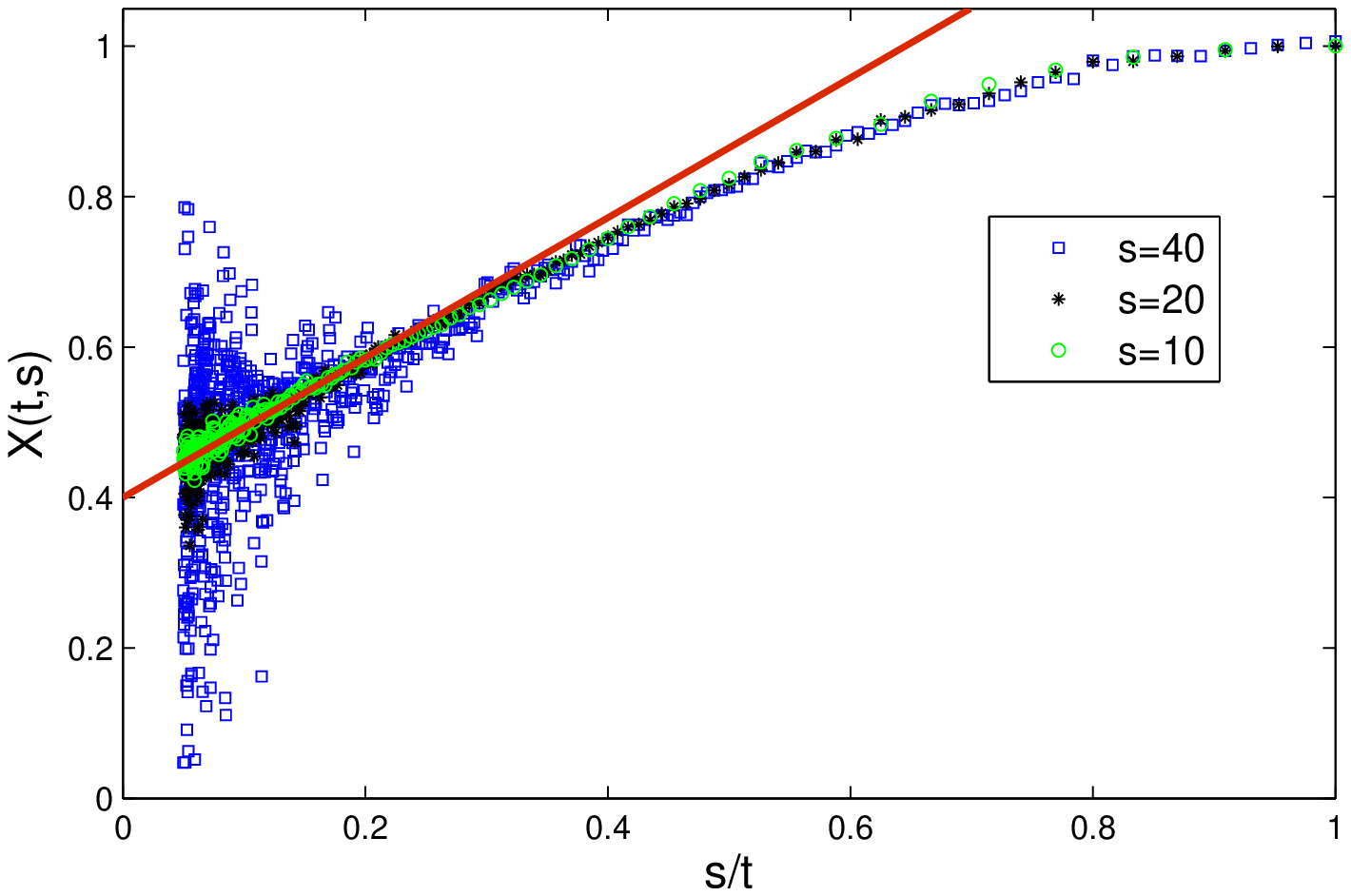}
\caption{(Color online.) Plot of $X(t,s)$ against $s/t$ for different fixed values of $s$ for the TFxTF (left) and SGXSG (right) quenched to $T_c$.}
\label{figx}
\end{centering}
\end{figure}

For the computation of the response function we have used the field-free 
algorithm introduced in \cite{lippiello05}. 
We have checked that the scaling
form (\ref{scalchi}) is obeyed with an exponent  consistent
with the expected behavior $a=b$. In Fig. \ref{figx}, we plot the 
fluctuation-dissipation ratio against $s/t$, for different values of $s$. The good data collapse observed confirms that $R$ scales as 
in Eq. (\ref{scalchi}), with $a=b$. For $s/t\simeq 1$, in 
the quasi-equilibrium regime, there
are no sensible deviations from the fluctuation-dissipation theorem, as discussed
in Sec. \ref{scalat}, and $X\simeq 1$. For $s/t \ll 1$ the so called aging regime
is accessed, ad $X$ lowers. Extrapolating the intercept at $s/t=0$ we obtain
$X_\infty =0.36\pm 0.03$ for the TFxTF and $X_\infty=0.40\pm 0.03$ for the SGxSG. 
Notice that these values are compatible within 
error bars. However, by comparing these values with the one ($X_\infty \simeq 0.33$) 
measured in the usual
square lattice, one sees that for the TFxTF they are only marginally compatible, 
while for the SGxSG they are totally incompatible.

In conclusion, in the case of critical quenches of the Ising model on the product
structures considered here one finds a finite $T_c$ and a pattern of behaviors
analogous to that found on homogeneous lattices. The value of the universal 
(on usual lattices) quantities
$a$ and $X_\infty$, appear to be different between the TFxTF, 
the SGxSG and the 2d lattice.

\section{Numerical results: Quench below $T_c$} \label{below}

\begin{figure}
\begin{centering}
\includegraphics[width=8.5 cm]{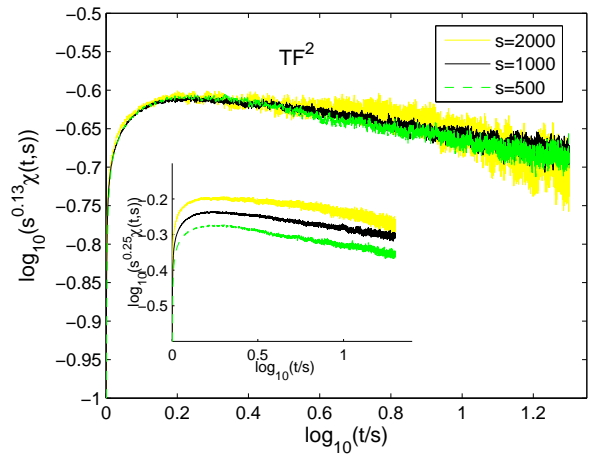}
\caption{(Color online.) $s^a\chi{}(t,s)$, with $a=0.13$, is plotted against $t/s$ for different fixed values of $s$ for the TFxTF quenched to $T=3.0<T_c$. In the inset we plot 
 $s^{0.25}\chi{}(t,s)$ against $t/s$ showing that for  $a=1/4$ the scaling is not satisfied.}

\label{figbelow}
\end{centering}
\end{figure}

For completeness, for the TFxTF we have computed the value of the exponent $a$ 
also in the case of a quench below $T_c$ ($X_\infty$ is trivially zero in this process,
see discussion at the end of Sec. \ref{scalbelow}). 
It must be recalled that in the Toblerone lattice
an exponent $a$ was found, whose value was compatible with the value $a=1/4$ found in the
two-dimensional case. This could suggest that the conjecture discussed above,
namely that products of {\it 1d-like} graphs could form a sort of universality class
sharing the exponent $a$,
although not confirmed by our previous data in critical quenches, 
could hold at least restricting to subcritical quenches. 
In order to check this issue we have computed
the integrated response function $\chi (t,s)=\int _s^t dt' R(t,t')$
and, recalling the additive form discussed in Sec. \ref{scalbelow}, 
we have isolated the aging term $\chi _{ag}$ by using a dynamics where
flips of spin in the bulk are forbidden, as discussed in \cite{nostriscalari,noi}.
From Eq. (\ref{scalchi}) one has 
\begin{equation}
\chi _{ag}(t,s)=s^{-a}\tilde h(t/s), 
\end{equation}
where $\tilde h(x)$ is another scaling function.
The data of Fig. (\ref{figbelow}) show that a good scaling collapse
is obtained with an exponent $a=0.13\pm.02$ which is different from the value
found in $d=2$ (see inset of Fig. (\ref{figbelow}) ). The fact that $a$ takes a comparable value in
the $d=2$ homogeneous lattice and on the Toblerone lattice, therefore, 
seems not to be a general property of
product graphs.

\section{Discussion and conclusions} \label{concl}

In this paper we have studied the scaling properties of the Ising
model quenched to or below the critical temperature
on graphs obtained by making direct products of {\it 1d-like} structures,
such as the TFxTF and the SGxSG. The direct product is  
a convenient tool to build graphs with a topology
sustaining a finite $T_c$, where
the critical properties
can be studied. This allows one to investigate the critical properties
of such structures and to determine the critical exponents.
Moreover, dynamical aspects can
also be considered, by studying the evolution
after temperature quenches. The aim of our analysis is to check the validity of
dynamical scaling on inhomogeneous
structures, and to study the behavior of the (limiting) fluctuation-dissipation ratios 
and of the scaling exponents.  On regular lattices these are well understood
and their  universal properties are well known. 
On product graphs we found 
a non-equilibrium scaling behavior similar to that
found on homogeneous lattices above $d_L$, where
time enters observable quantities
through a single growing length representing the size of
the critical correlations established.  
Regarding the quantities which on homogeneous lattices are universal, previous 
studies of phase-ordering on a certain class of graphs
showed their dependence on several parameters, among which
the temperature, at variance with the regularity observed on homogeneous
lattices. A notable exception was represented by the exponent
$a$ and the limiting fluctuation-dissipation ratio $X_\infty$,
which were found to take the same values $a=0$ and $X_\infty=1/2$ on all
the graphs with $T_c=0$ considered insofar. 
In this paper we have considered the possibility that such a regularity
could be extended to the various direct products of these $T_c=0$ graphs than one can consider,
by studying if critical exponents or $X_\infty$ take a unique
value for all the graphs of this class. 
The results we found, however, are negative in this respect, either in critical quenches
or in sub-critical quenches. Indeed we have shown that 
product graphs exhibit different
exponents, including $a$, and, restricting to critical
quenches, also a different value of $X_\infty$. 
This may indicate that a robust universality property
as observed on usual lattices is lost in the
realm of inhomogeneous structure, or that, if some
analogue of it exists, the relevant parameters of the
graph topology determining universality have not
yet been identified. 

\section*{Acknowledgments}

F.Corberi acknowledges financial support
from PRIN 2007 JHLPEZ ({\it Statistical Physics of Strongly Correlated
Systems in Equilibrium and Out of Equilibrium: Exact Results and 
Field Theory Methods}).

\section*{Appendix}

We define the {\it quality} of the collapse between the different curves as follows.
First, for every time $t_i$  we fit the curve
$\log(F_{G_k}(t_i))$ as a function of the size $R_k$ with a polynomial of the 
form $\tilde F(t_i,x)$ where $x=\log (R_k/t_i^{1/z})$. This is done since the function 
$F_{G_k}(t)$ is known only on a discrete set of sizes $R_k$.
The curves $\hat F(x,t_i)=\log (t_i^{b-d_f/z})+\tilde F(t_i,x)$, obtained
at different times $t_i$ should collapse once plotted against $x$  
(the reason for taking the logarithms is
that, since one has power-law dependences, the logarithms allow one to better
take into account deviations from perfect collapse on a wide range of sizes 
and times).
Then we introduce a {\it squared distance} between the rescaled curves
as
\begin{equation}
S\left(T,z,b\right)=\sum_{i} \int \left[\hat F(x,t_i)-\hat F(x,t_{i+1})\right ]^2dx
\end{equation}


\begin{thebibliography}{99}
\bibitem{nota}
This property was not shown in \cite{nostriscalari}.
\end{thebibliography}
\end{document}